\documentclass{jpsj3}
\usepackage{txfonts}
\usepackage{txfonts}
\usepackage{bm} 
\usepackage{amsmath} 
\usepackage{mathrsfs}
\usepackage{array}
\usepackage[dvipdfmx]{graphicx}
\usepackage{algorithm}
\usepackage{algorithmic}
\usepackage{comment}
\title{Deep Neural Network Detects Quantum Phase Transition}

\author{Shunta Arai\thanks{arai@smapip.is.tohoku.ac.jp}, Masayuki Ohzeki and Kazuyuki Tanaka}
\inst{Graduate School of Information Sciences, Tohoku University, Sendai 980-8579, Japan}

\abst{We detect the quantum phase transition of a quantum many-body system by mapping the observed results of the quantum state onto a neural network. 
In the present study, we utilized the simplest case of a quantum many-body system, namely a one-dimensional chain of Ising spins with the transverse Ising model.
We prepared several spin configurations, which were obtained using repeated observations of the model for a particular strength of the transverse field, as input data for the neural network. 
Although the proposed method can be employed using experimental observations of quantum many-body systems, we tested our technique with spin configurations generated by a quantum Monte Carlo simulation without initial relaxation. 
The neural network successfully classified the strength of transverse field only from the spin configurations, leading to consistent estimations of the critical point of our model $\Gamma_c =J$. }

\begin{document}
\maketitle

In recent years, the technique of deep learning has become well established and has shown outstanding  performance in various fields. 
For instance, GoogLeNets is  a deep neural network (DNN) that won a prize at the ImageNet Large Scale Visual Recognition Competition in 2014 (ILSVRC14) \cite{googlenets}.
AlphaGo is a DNN designed by Deep Mind that is trained using a novel combination of supervised learning from human expert games and reinforcement learning from games of self play \cite{alphago}. 
AlphaGo has successfully beat many human Go players. 

In general, DNNs can be used extract unclear features of a given input dataset or to identify hidden relationships between input and output.
DNNs have been applied to various problems in physics \cite{DL_ising,deep_2dspin,ml_phase}.
In addition, a theoretical assessment of a DNN has been performed using several tools in the theoretical physics \cite{Ohzeki2015}.
In the previous study performed by Tanaka and Tomita \cite{detection_cnn}, they have utilized convolutional neural networks (CNN) for detecting the phase transition of a classical two-dimensional Ising model on a square lattice using only the spin configurations obtained by the Markov-chain Monte Carlo method. 
The sharp change of weights in the resulting CNN signified the phase transition and successfully computed a reasonable value of the critical point of the two-dimensional Ising model on a square lattice.
This neural network (NN) approach is not restricted to the case of the classical Ising model. 
In general, NNs have the capability of being applied to various types of data.
Thus, in the present study, we applied the technique established in the previous study to a quantum many-body system.
In a quantum many-body system, we can observe the microscopic state through measurement.
To obtain the expectation value of a physical observable, the measurements must be repeated.
In particular, one of the typical measurements of physical observables in the quantum spin systems is the direction of the spin variables.

In the present study, we consider the one-dimensional transverse-field Ising model, which shows a quantum phase transition.
The present study is intended to serve as a test study for the use of NNs in quantum many-body systems.
We also investigate the significance of the NN structure.
Unlike the previous study \cite{detection_cnn}, we employ multi-layer perceptrons (MLPs), which are the simplest form of a NN.
Despite their simplicity, MLPs can be used to provide a robust NN. In particular, CNNs can capture complex features on images by introducing the convolution process to MLPs.
This process functions as a kind of real-space renormalization-group analysis \cite{DeepRG}, which enables elucidation of the critical behavior of the system \cite{Ohzeki2009}.
However, through our analysis, we confirmed that the process of the convolution does not necessarily affect the extraction of critical behaviors.
The simple MLP can also be used to estimate the precise value of critical points.

We consider a one-dimensional transverse-field Ising model, defined by the following Hamiltonian: 
\begin{align}
\mathscr{H} =\mathscr{H}_{0}+\mathscr{H}_{1},
\end{align}
where we define
\begin{equation}
\mathscr{H}_{0}=  -J\sum_{i=1}^{L} \sigma_{i}^{z} \sigma_{i+1}^{z} -h\sum_{i=1}^L \sigma_{i}^{z},
\end{equation}
and
\begin{equation}
\mathscr{H}_{1}=- \Gamma \sum_{i=1}^{L} \sigma_{i}^{x}.
\end{equation}
Here, $h$ stands for the strength of the longitudinal  magnetic field, $\Gamma$ represents the strength of the transverse magnetic field, $\sigma_{i}^{z}$ is the $z$ component of the Pauli matrix at site $i$, and $\sigma_{i}^{x}$ is the $x$ component of the Pauli matrix at site $i$. 
The symbol $L$ is the number of spins in the one-dimensional chain.
This model has a quantum phase transition at $\Gamma_c =J$. 
The ordered phase occurs when $J>\Gamma$, and the disordered phase occurs when $\Gamma >J$. 
However, we cannot directly simulate the quantum many-body system because the Hamiltonian includes non-commuting operators.
We then employ the Suzuki--Trotter decomposition\cite{suzuki_trotter} to express our model in terms of the c numbers:
\begin{align}
\exp(-\beta \mathscr{H})=\lim_{\tau \to \infty}\left(\exp\left(-\frac{\beta}{\tau}\mathscr{H}_{0}\right) \exp\left(-\frac{\beta}{\tau}\mathscr{H}_{1}\right)\right) ^{\tau},\label{suzukiT}
\end{align}
where $\beta$ is the inverse temperature and $\beta = 1/{T}$. 
The Trotter number is defined as $\tau$.
The one-dimensional transverse-field Ising model is then mapped onto a two-dimensional classical Ising model using Eq. \eqref{suzukiT}. 
The effective Hamiltonian is given by  
 \begin{align}
\mathscr{H}_{\rm{eff}}  =& - \frac{J}{\tau} \sum_{t=1}^{\tau} \sum_{i=1}^{L} \sigma_{it} \sigma_{i+1 t} - \frac{h}{\tau}\sum_{t=1}^{\tau}\sum_{i=1}^{L} \sigma_{it} - \gamma \sum_{t=1}^{\tau} \sum_{i=1}^{L} \sigma_{it} \sigma_{i  t+1},
\end{align} 
where $\gamma=-\log(\tanh(\beta \Gamma/{\tau}))/ 2\beta$ and $\sigma_{i,t}=\{ -1,1\}$. 
We impose the periodic boundary conditions $\sigma_{L+1,t}=\sigma_{1,t} $ and $\sigma_{i,\tau+1}=\sigma_{i,1}$.
Therefore, our model becomes a classical model.
In the classical expression, one dimension is the original spatial dimension, while the other dimension expresses imaginary time.
Imaginary time can be interpreted as repeated observations on the quantum system.
We regard the two-dimensional expression of our model as a sequence of measurement results of spin configurations on the one-dimensional chain.
The following analysis is not restricted to the case of the one-dimensional Ising model with a transverse field.
In general, our analysis can be applied to any quantum many-body system.
In the present study, we demonstrate the procedure for detecting the quantum phase transition and reduce the complexity in constructing the NN to simplify this process, making it more accessible than the previous study.

The technique can be applied to the so-called non-stochastic Hamiltonian, on which quantum Monte Carlo simulation cannot be applied in a straightforward manner\cite{Bravyi2008}.
Therefore, the use of classical computer simulation for obtaining the microscopic state is not useful except for several special cases \cite{Ohzeki2017}.
In order to obtain the spin configurations, we may also utilize a quantum simulator such as the D-Wave machine, which is a well-known system that has performed manipulation of a $2000$-bit Ising model using quantum annealing by tuning the strength of the transverse-field on the chimera graph \cite{Dwave2010a,Dwave2010b,Dwave2010c}.

Here, we show the process of training the MLP in detail for those who may be unfamiliar with machine learning. 
The MLP utilizes a complicated function to relate input to output by repeated linear and nonlinear transformations.
We prepare the pairs of input and output data and fit the MLP to these data, while we optimize the  structure of the MLP.
This technique is known as supervised learning \cite{prml}.
We set the discretized values of the transverse fields as the input and the spin configurations as the output:
\begin{align}
D_{\rm{Transverse}} = \{ \bm{\sigma^{d}} ,{\Gamma}^{d} \}_{d=1,2,...,D} , \label{qmcdata}
 \end{align} 
where  the spin configurations $\bm{\sigma^{d}}$ are denoted
\begin{equation}
\bm {\sigma}^d =\begin{pmatrix}
	\sigma_{11}^d & \dots & \sigma_{1\tau}^d \\
	\vdots &  \ddots & \vdots			      \\
	\sigma_{L1}^d & \dots & \sigma_{ L\tau}^d
	\end{pmatrix}.
	\label{latticetrans}
\end{equation}
The first subscript denotes the index of the space and the second subscript denotes the index of imaginary time in the quantum Monte-Carlo simulation or that of the sequence of repeated measurements from an actual experiment.

To utilize the output data in training the MLP, we employ the following flattened representation of $\bm{\sigma}^d$ as 
\begin{align}
 \vec{\sigma^d} =f_{\rm{Flatten}} (\bm{\sigma}^d) =  (\sigma_{11}^d,\sigma_{21}^d,...,\sigma_{L1}^d,\sigma_{12}^d,...,\sigma_{L \tau}^d).
 \label{flattrans}
\end{align}
Here, $\tau$ is the Trotter number.
Detecting the value of the transverse field from the spin configurations is considered a multi-classification problem. 
We discretize real-value $\Gamma\in [0,2)$ to $\vec{\Gamma}$ by using the function
\begin{align}
\Vec{x}={f}_{\rm{class}}(x) =\begin{cases}
(1,0,...,0,0) & \text{for $x \in [0,\frac{2}{N_{\rm{class}}})$} \\
(0,1,...,0,0)  &\text{for $x \in [\frac{2}{N_{\rm{class}}},\frac{4}{N_{\rm{class}}})$} \\
~~~~~~~~~... \\
(0,0,...,1,0) & \text{for $x \in [ 2\times \frac{N_{\rm{class}}-2}{N_{\rm{class}}},2\times \frac{N_{\rm{class}}-1}{N_{\rm{class}}})$} \\
(0,0,...,0,1)  & \text{for $x \in [ 2\times \frac{N_{\rm{class}}-1}{N_{\rm{class}}},2)$} \\ 
\end{cases},
\label{discretized_func}
\end{align}
where $N_{\rm{class}}$ is the number of classes. 
This discretization is often called the one-hot representation.  

Unlike the previous study\cite{detection_cnn}, we consider a very simple three-layer MLP in the present study, consisting of the 
input, hidden, and output layers. 
We stack these layers to design a nontrivial function. 

From the input layer to the hidden layer, we compute
\begin{align}
{u}_k^{(1)}&=\bm{W}_{k}^{(1)} \vec{\sigma}^{} +{b}^{(1)}\label{fu}\\
z_k^{(1)} &= f_{\rm{ReLU}}(u_k^{(1)})
\end{align}
for $k=1,2,...,N_h$, where $W_{k}^{(1)}$ is the weight parameter connecting to unit $k$ in the hidden layer, $N_h$ is the number of hidden units, and $b^{(1)}$ represents the bias in the input layer.
The weights and bias perform the linear transformation.
We use the rectified linear function (ReLU) \cite{relu} as the nonlinear transformation, which is often called the activation function. 
The ReLU function is defined by
\begin{align}
f_{\rm{ReLU}} (x) =\max(0,x)\label{reluf}.
\end{align}
This type of activation function is employed to prevent the gradient from vanishing, which hampers efficient learning by use of the gradient descent method.
From the hidden layer to the output layer, we compute 
\begin{align}
{u}_k^{(2)}&=\bm{W}_{k}^{(2)} \vec{z}^{(1)} +{b}^{(2)}\label{fu}\\
y_k^{(2)} &=f_{\rm{softmax}}(u_{k}^{(2)})= \frac{\exp(u_k^{(2)})}{\sum_{k=1}^{N_{\rm{class}}} \exp{(u_k^{(2)})}} \label{softmaxf}
\end{align}
for $k=1,2,...,N_{\rm{class}}$. 
We represent $(z_1,z_2,...,z_{N_h})^{\rm{T}}$ as  $\vec{z}^{(1)}$. 
The activation function \eqref{softmaxf} is called the softmax function. 
These processes are are known as $forward$ $propagation$. 
We then calculate the loss function, which is given by  
\begin{align}
E(\vec{y},\vec{\Gamma})=-\sum_{k=1}^{N_{\rm{class}}}\Gamma_k \log (y_k) ,
\label{cross_entropy}
\end{align}
where $\Gamma_k$ is the $k$th element of $\vec{\Gamma}$.
This loss function is called the cross entropy. 
We optimize the summation of the loss function over all components of the dataset and update parameters $W, b$ using gradient descent.
One efficient algorithm for optimizing the weights while avoiding a learning plateau at the saddle point of the loss function is the Adam method\cite{adam}.
To obtain generalized performance, we employ mini-batch learning\cite{minibatch} in which the summation of the loss function is approximated by a partial summation over the mini-batch of randomly chosen components from the dataset.
When the batch size is small, the MLP tends to attain generalized performance \cite{GE2017}.
We utilize the weight decay \cite{weight_decay} via $L_2$-norm regularization to avoid overfitting to the training data.
This technique is a standard method for achieving more generalized MLP performance.

The process of calculating the loss function and updating the weights is called $back$  $propagation$. 
We show the MLP learning protocol in Algorithm \ref{alg1}. The epoch is the step of learning. 
The symbol $N_m$ represents the mini-batch number.
\begin{algorithm}                      
\caption{Learning protocol  for MLP}                                
\begin{algorithmic}      
\STATE Input: $D_{\rm{Transverse}},  W, b ,N_{{h}},N_{\rm{class}}, N_{{m}}$,epoch   
\STATE  Set Parameters  $W, b$ epoch
\FOR{$i=1$ to epoch} 	
\FORALL{$D_{\rm{Transverse}}$}
\STATE  1: Randomly pick up $N_{\rm{m}}$ data in $D_{\rm{Transverse}}$
\STATE  2: Compute the output \eqref{softmaxf} for $k=1,2,...,N_{\rm{class}}$\\(forward propagation) 
\STATE  3: Calculate the loss function \eqref{cross_entropy} and update parameters\\ using Adam \\(back propagation)
\ENDFOR
\ENDFOR
\end{algorithmic}
\label{alg1}
\end{algorithm}
$\\$
$\\$

Before discussing our experimental results, we describe our experimental environment. 
We perform the quantum Monte Carlo simulation\cite{quantum_monte} (MCS) of the effective Hamiltonian \eqref{qmcdata}, while tuning the transverse field from its maximum $ \Gamma_{\rm{max}}$ to minimum $\Gamma_{\rm{min}}$ by $\delta{\Gamma}=2/N_{\rm{class}}$.
We then set the parameters $J=1, \Gamma_{\rm{max}}=1.99,  \Gamma_{\rm{min}}=0.01 , D=10^4,
\tau=40, \beta=50, L=40$, and $h=-10^{-3}$.
We generate $100$ spin configurations $\bm{\sigma}$ for each class $k$, and 
we take $100$ MCS for each sample using the Metropolis method \cite{monte_carlo_sp,binder_monte}.
We do not discard spin configurations generated in the initial relaxation. 
Starting from the random initial configurations, we obtain $100$ spin configurations at each $k$.
Therefore, the spin configurations are not necessarily in equilibrium around the critical point.
In this sense, MLP might be able to infer the strength of the transverse field from non-equilibrium behavior such as the nonequilibrium relaxation \cite{NE1998}. 
In this regard, we do not have any definite conclusion about the effectiveness of the nonequilibrium behavior.
However, when the equilibrium spin configurations, which are the output from the system, result in similar behavior of the weight in the MLP, we cannot obtain the precise estimation as shown below.
We will discuss this point in more detail later.
 
In the present three-layer MLP, 
the number of units in the input layer is $L \times \tau$, the number of units in the hidden layer is $N_h=2000$, and 
the number of units in the output layer is $N_{\rm{class}}$. 
According to Algorithm \ref{alg1}, MLP learns the dataset $D_{\rm{Transverse}}$. 
We show the heat map of weights from the hidden layer to the output layer after $50$ learning epochs in Fig.2. 
We observe a sharp change in the weights near $\Gamma_{c}$. Figure 2 indicates that the distribution of weights changes near the critical point $\Gamma_{c}=1$. 

Similarly to the previous study\cite{detection_cnn}, to extract a characteristic point for weights, we define 
\begin{align}
W_{\rm{mean}}(\Bar{\Gamma}(k)) &=  \frac{1}{|\bm{W}_{k}^{(2)}|} \bm{1}^{\rm{T}} \bm{W}_{k} ^{(2)},\label{order_para_t1}
\end{align}
where $k$ represents the class of the discretized transverse field and $\bm{W}_{k}^{(2)}$ is the weight connected to unit $k$ in output layer. 
The weight can be interpreted as a type of magnetization order parameter.

The notation $|\bm{W}_{k}^{(2)}|$ indicates the number of $\bm{W}_{k}^{(2)}$. 
The vector $\bm{1}^T$ is defined by $\bm{1}^{T}=(1,1,...,1)$ and has $N_h$ components. 
In the present study, we introduce the variance of $W_k^{(2)}$ as 
\begin{align}
{W}_{\rm{var}}(\Bar{\Gamma}(k)) &= \frac{1}{|\bm{W}_{k}^{(2)}|}(\bm{W}_{k}^{(2)}-W_{\rm{mean}} \bm{1})^{\rm{T}}(\bm{W}_{k}^{(2)}-W_{\rm{mean}}\bm{1}).\label{order_para_t2}
\end{align}
This is the variance of the order parameters, which can often signify the phase transition in the Markov Chain Monte Carlo method by computing the binder parameter for estimating the precise critical point location.
We substitute $\Bar{\Gamma}(k)$ for class $k$ as 
\begin{align}
\Bar{\Gamma}(k)=\left({\frac{2k-2}{N_{\rm{class}}}+\frac{2k}{N_{\rm{class}}}}\right) \times \frac{1}{2}=\frac{2k-1}{N_{\rm{class}}}, \label{orderg}
\end{align} 
where $\Bar{\Gamma}(k)$ represents a transverse field corresponding to a certain class $k$. 
\begin{figure}[t]
\begin{center}
\includegraphics[width=8cm]{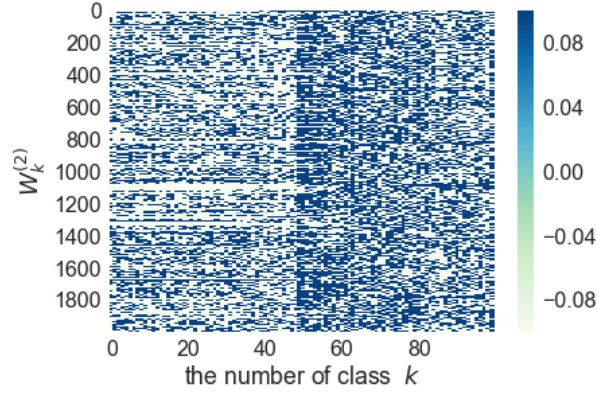} 
\caption{(Color online) Heat map of MLP from the hidden layer to the output layer. 
The horizontal axis  represents class $k$, which corresponds to the transverse field \eqref{orderg}. 
The vertical axis illustrates the weight of the MLP, which is connected to unit $k$ in the output layer.
}
\end{center}
\label{heatmap_mlp} 
\end{figure}
\begin{figure}[t]
\begin{center}
\includegraphics[width=85mm]{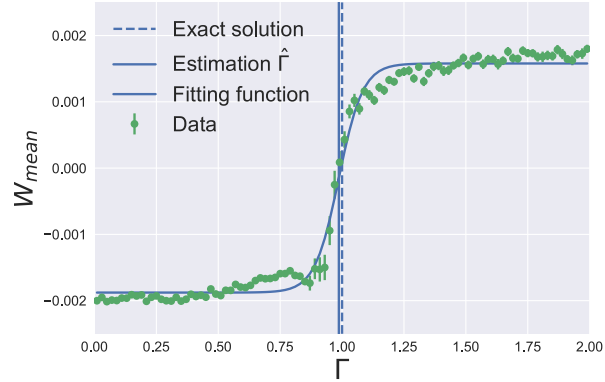} 
\end{center}
\caption{(Color online) Quantity \eqref{order_para_t1} and fitting curve \eqref{tanhg}. 
The green point represents the quantity \eqref{order_para_t1}. 
The horizontal axis is the transverse field \eqref{orderg}, while  
the vertical axis denotes the quantity \eqref{order_para_t1} at each transverse field. 
The blue straight line is the location of the predicted solution given by \eqref{tanhg}, while 
the blue dotted line is the exact solution of the transverse-field Ising model $\Gamma_c=J$. 
The error bar is given by the standard error among $100$ datasets $D_{\rm{Transverse}}$}$ $
\label{meanw_trans} 
\end{figure}

\begin{figure}[t]
\begin{center}
\includegraphics[width=75mm]{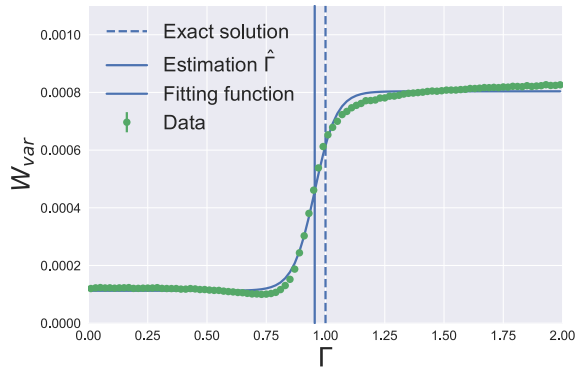} 
\end{center}
\caption{(Color online) Quantity \eqref{order_para_t2} and fitting curve \eqref{tanhg}. 
The green points are the quantity \eqref{order_para_t2}. 
The vertical axis denotes the quantity \eqref{order_para_t2} at each transverse field.
The horizontal axis and curves are the same as those used in Figure \ref{meanw_trans}.}
\label{varw_trans} 
\end{figure}

Similarly to the previous research \cite{detection_cnn}, we fit two of the defined quantities to the following function using the non-linear least square method:
\begin{align}
\hat{W} (\Bar{\Gamma}(k)) = a\tanh b(\Bar{\Gamma} - \hat{\Gamma})+c \label{tanhg},
\end{align}
where  $a,b, \hat{\Gamma},c$ are parameters, and $\hat{\Gamma}$ is the estimation of the critical point. 
The point $\hat{\Gamma}$ indicates the location where this function sharply changes, as shown in Figs. \ref{meanw_trans} and \ref{varw_trans}.
We obtain the estimation of $\hat{\Gamma}=0.9870$ from \eqref{order_para_t1} and $\hat{\Gamma}=0.9544$ from \eqref{order_para_t2}. 
The error rate between the exact solution $\Gamma_c=1$ and the $\hat{\Gamma}$ estimation from \eqref{order_para_t1} and \eqref{order_para_t2} is $1.3\%$ and $4.56\%$, respectively. 
The order parameters \eqref{order_para_t1} and \eqref{order_para_t2} change slowly, unlike those in Figs. \ref{meanw_trans} and \ref{varw_trans} when we take $10000$ MCS to equilibrate the system. 
Therefore, the MLP can determine the change of phase from the correlations between spin configurations of different transverse field values.
In other words, MLP captures the nonequilibrium behavior of the system.
However, the weight of the MLP shows different distributions, even for equilibrium spin configurations.
Thus, increasing the number of data $D$ may lead to a better estimation.
This is still a remaining problem.

In conclusion, we have shown that the MLP can be used to detect quantum phase transition by observing changes in the weights. 
The results show fairly good estimations of the critical point location in the transverse-field Ising model.
The technique used in the present study can be applied to sequential observations of experimental results instead of the imaginary-time distribution of the quantum systems. 
Typically, the Monte Carlo simulation requires a lot of computational time to precisely estimate the location of the critical point because of the number of relaxation times required to obtain an equilibrium distribution. 
On the other hand, our estimation is accurate even without long relaxation times.
In addition, unlike the previous study \cite{detection_cnn}, which tested the CNN on the two-dimensional Ising model on a square lattice, we utilized simple MLPs in this study.
We showed that the one-dimensional transverse-field Ising model is essentially the same as the classical two-dimensional Ising model.
We further confirmed that the MLP has sufficient performance to detect critical behavior. 
Therefore, it is not necessary to utilize the convolutional process to detect critical behavior. 
We emphasize that this method has the potential to be applied to models with non-trivial order parameters. 
In our future work, we plan to expand this method to models with topological phases such as XY models.

\section*{Acknowledgments}
We acknowledge the help of Shun Kataoka and Yuya Seki for many fruitful discussions about the numerical simulation. 
M.O. was supported by KAKENHI No.15H0369, No.16H04382, and No.16K13849 and JST-START. 
This work was partly supported by JST-CREST (No.JPMJCR1402). 
\bibliography{main}
\end{document}